\documentclass{aa}
\usepackage{graphicx,natbib,epsfig}

\begin{document}
\title{Modeling the Jovian subnebula: II - Composition of regular satellites ices}

\author{Olivier Mousis \inst{1}\fnmsep \inst{2} \and Yann Alibert \inst{2}
          }

   \offprints{O.Mousis}

   \institute{Observatoire de Besan\c{c}on, CNRS-UMR 6091, BP 1615, 25010
Besan\c{c}on Cedex, France
         \and
            Physikalisches Institut, University of Bern, Sidlerstrasse 5,
CH-3012 Bern, Switzerland \\
             \email{Olivier.Mousis@obs-besancon.fr}\\
             }

\titlerunning{Composition of regular satellites ices}

\abstract{
We use the evolutionary turbulent model of Jupiter's subnebula described by Alibert et al. (2005a) to constrain the composition of ices incorporated in its regular icy satellites. We consider CO$_2$, CO, CH$_4$, N$_2$, NH$_3$, H$_2$S, Ar, Kr, and Xe as the major volatile species existing in the gas-phase of the solar nebula. All these volatile species, except CO$_2$ which crystallized as a pure condensate, are assumed to be trapped by H$_2$O to form hydrates or clathrate hydrates in the solar nebula. Once condensed, these ices were incorporated into the growing planetesimals produced in the feeding zone of proto-Jupiter. Some of these solids then flowed from the solar nebula to the subnebula, and may have been accreted by the forming Jovian regular satellites. We show that ices embedded in solids entering at early epochs into the Jovian subdisk were all vaporized. This leads us to consider two different scenarios of regular icy satellites formation in order to estimate the composition of the ices they contain. In the first scenario, icy satellites were accreted from planetesimals that have been produced in Jupiter's feeding zone without further vaporization, whereas, in the second scenario, icy satellites were accreted from planetesimals produced in the Jovian subnebula. In this latter case, we study the evolution of carbon and nitrogen gas-phase chemistries in the Jovian subnebula and we show that the conversions of N$_2$ to NH$_3$, of CO to CO$_2$, and of CO to CH$_4$ were all inhibited in the major part of the subdisk.  Finally, we assess the mass abundances of the major volatile species with respect to H$_2$O in the interiors of the Jovian regular icy satellites. Our results are then compatible with the detection of CO$_2$ on the surfaces of Callisto and Ganymede and with the presence of NH$_3$ envisaged in subsurface oceans within Ganymede and Callisto.
   }

\maketitle

\section{Introduction}
 In a recent paper, Alibert et al. (2005a) (hereafter referred to as Paper I) developed a two-dimensional time dependent $\alpha$-turbulent model of the Jovian subnebula whose evolution is ruled by the last sequence of Jupiter formation. These authors carried out migration calculations in the Jovian subnebula in order to follow the evolution of the ices/rocks ratios in the protosatellites as a function of their migration pathways. By tempting to reproduce the distance distribution of the Galilean satellites, as well as their ices/rocks ratios, they obtained some constraints on the viscosity parameter of the Jovian subnebula and on its thermodynamical conditions.
 
 They showed that the Jovian subnebula evolves in two distinct phases during its lifetime. In the first phase, the subnebula is fed through its outer edge by gas and gas-coupled solids originating from the protoplanetary disk as long as it has not been dissipated. During the major part of this period, temperature and pressure conditions in the Jovian subnebula are high enough to vaporize any icy planetesimal coming through. When the solar nebula has disappeared, the subnebula enters the second phase of its evolution. The mass flux at the outer edge stops, and the Jovian subnebula gradually empties by accreting its material onto the forming Jupiter. At the same time, due to angular momentum conservation, the subnebula expands outward. Such an evolution implies a rapid decrease of temperature, pressure and surface density conditions over several orders of magnitude in the whole Jovian subnebula. 

In the present work we focus on the possibility of estimating the composition of ices incorporated in the regular icy satellites of Jupiter in the framework of the model described in Paper I. A similar study was previously conducted by Mousis \& Gautier (2004) (hereafter referred to as MG04) but here we present several significant improvements.

First, the initial accretion rate of our turbulent model of the Jovian subnebula is fully consistent with that calculated in the last phase of Jupiter formation (see Paper I). As a result, the temporal evolution of our model of the Jovian subnebula, as well as the thermodynamical conditions inside the subnebula, are quite different from that of MG04. Hence, the question of the resulting composition of ices incorporated in the Galilean satellites remains open. 

Second, in our model, the solids flowing in the subnebula from the nebula were formed in Jupiter's feeding zone. For the sake of consistency, it is important to calculate their composition using the same thermodynamical and gas-phase conditions as those considered by Alibert et al. (2005b - hereafter referred to as A05b). Indeed, using the clathrate hydrate trapping theory (Lunine \& Stevenson 1985), A05b have interpreted the volatile enrichments in Jupiter's atmosphere, in a way compatible with internal structure models derived by Saumon and Guillot (2004). As a result, they determined the range of valid CO$_2$:CO:CH$_4$ and N$_2$:NH$_3$ gas-phase ratios in the solar nebula to explain the measured enrichments, and the minimum H$_2$O/H$_2$ gas-phase ratio required to trap the different volatile species as clathrate hydrates or hydrates in icy solids produced in Jupiter's feeding zone.
Our calculations then allow us to determine the composition of ices in Jupiter's regular satellites, in a way consistent with the enrichments in volatile species observed in the giant planet's atmosphere by the
{\it Galileo} probe.

Finally, we consider further volatile species that are likely to exist in the interiors  of the Jovian regular icy satellites. In addition to CO, CH$_4$, N$_2$, and NH$_3$ that have already been taken into account in the calculations of MG04, we also consider CO$_2$, Ar, Kr, Xe and H$_2$S. CO$_2$ has been detected on the surface of Ganymede and Callisto (McCord et al. 1998; Hibbitts et al. 2000, 2002, 2003) and is likely to be a major carbon compound in the initial gas-phase of the solar nebula since large quantities are observed in the ISM (Gibb et al. 2004). Moreover, Ar, Kr, Xe and H$_2$S abundances have been measured in the atmosphere of Jupiter (Owen et al. 1999). Since, according to A05b, these volatile species have been trapped  in icy planetesimals in Jupiter's feeding zone during its formation, they may also have been incorporated into the material (gas and solids) delivered by the solar nebula to the Jovian subnebula and taking part in the formation of the regular satellites.

The outline of the paper is as follows. In Sect. 2, we examine the conditions of volatiles trapping in solids formed in Jupiter's feeding zone. In Sect. 3, we recall some details of the thermodynamical characteristics of our turbulent model of the Jovian subnebula. This allows us to investigate the conditions of survival of these solids formed inside the solar nebula and accreted by the subnebula. In this Section, we also study the evolution of the gas-phase chemistries of carbon and nitrogen volatile species in the subdisk. In Sect. 4, we estimate the mass ratios with respect to water of the considered volatile species in the interiors of regular icy satellites. Sect. 5 is devoted to discussion and summary.

\section{Trapping volatiles in planetesimals formed in Jupiter's feeding zone}
The volatiles ultimately  incorporated in the regular icy satellites have been first trapped under the form of hydrates, clathrate hydrates or pure condensates in Jupiter's feeding zone. The clathration and hydratation processes result from the presence of crystalline water ice at the time of volatiles trapping in the solar nebula. This latter statement is justified by current scenarios of the formation of the solar nebula  who consider that most of ices falling from the presolar cloud onto the disk vaporized when entering in the early nebula. Following Chick and Cassen (1997), H$_2$O ice vaporized within 30 AU in the solar nebula. With time, the decrease of temperature and pressure conditions allowed the water to condense and form microscopic crystalline ices (Kouchi et al. 1994, Mousis et al. 2000). Once formed, the different ices agglomerated and were incorporated into the growing planetesimals. These planetesimals may ultimately have been part of the material (gas and solid) flowing in the subnebula from the solar nebula. Moreover, larger planetesimals, with metric to kilometric dimensions, may have been captured by the Jovian subnebula when they came through.

In the model we consider, Jupiter forms from an embryo initially located at $\sim$ 9-10 AU (Alibert et al. 2005c). Since the subnebula appears only during the late stage of Jupiter formation when the planet has nearly reached its present day location (see Paper I for details), we used the solar nebula thermodynamical conditions at 5 AU. On the other hand, the use of the solar nebula thermodynamical conditions at $\sim 10$ AU would not change our conclusions, since the composition of icy planetesimals does not vary significantly along the migration path of Jupiter if a similar gas-phase composition is assumed (A05b).

\begin{figure}
   \begin{center}
   \epsfig{file=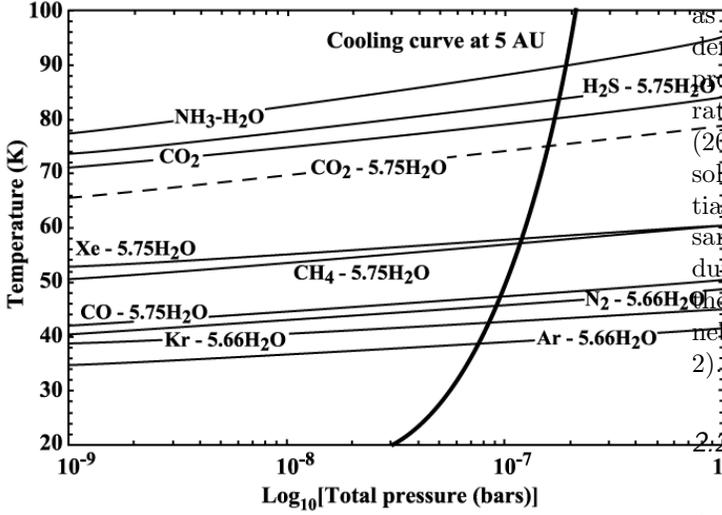,angle=-90,width=100mm}
   \end{center}
   \caption{Stability curves of the species trapped as hydrates or clathrate hydrates considered in this work and evolutionary track of the nebula in $P-T$ space at the heliocentric distance of 5 AU. Abundances of various elements are solar. For CO$_2$, CO and CH$_4$, their abundances are calculated assuming CO$_2$:CO:CH$_4$~=~30:10:1. For N$_2$ and NH$_3$, their abundances are calculated assuming N$_2$:NH$_3$~=~1. The condensation curve of CO$_2$ pure condensate (solid line) is plotted together with that of the corresponding clathrate hydrate (dashed line). The solar nebula cooling curve at 5 AU is derived from A05b.}
   \label{cool_curve}
\end{figure}

The trapping process of volatiles, illustrated in Fig. \ref{cool_curve}, is calculated using the stability curves of clathrate hydrates derived from the thermodynamical data of Lunine \& Stevenson (1985) and the cooling curve at 5 AU taken from the solar nebula model used to calculated Jupiter's formation (see A05b). For each considered ice, the domain of stability is the region located below its corresponding stability curve. Note that the use of cooling curves derived from others evolutionary $\alpha$-turbulent models of the solar nebula (the nominal models of Drouart et al. (1999) and Hersant et al. (2001)) intercept the stability curves of the different condensates at similar temperature and pressure conditions.
 The stability curve of CO$_2$ pure condensate is derived from the existing experimental data (Lide 1999). From Fig. \ref{cool_curve}, it can be seen that CO$_2$ crystallizes as a pure condensate prior to being trapped by water to form a clathrate hydrate during the cooling of the solar nebula. Hence, we assume in this work that solid CO$_2$ is the only existing condensed form of CO$_2$ in the solar nebula.

\subsection{Initial ratios of CO$_2$:CO:CH$_4$ and N$_2$:NH$_3$ in the solar
nebula gas-phase}

In the present work, the abundances of all elements are considered to be solar (Anders \& Grevesse 1989) and O, C, and N exist only under the form of H$_2$O, CO$_2$, CO, CH$_4$, N$_2$, and NH$_3$ in the solar nebula vapor phase. Gas-phase abundances relative to H$_2$ in the nebula for species considered here are given in Table \ref{table_AG89}.

\begin{table}[h]
   \caption[]{Gas phase abundances of major species with respect to H$_2$ in the
solar nebula 
(from Anders \& Grevesse 1989) for CO$_2$:CO:CH$_4$~=~30:10:1 and
N$_2$:NH$_3$~=~1.}
  \begin{center}
       \begin{tabular}[]{lclc}
            \hline
            \hline
            \noalign{\smallskip}
             Species $i$ & $x_i$ & Species $i$ & $x_i$\\
            \noalign{\smallskip}
             \hline
             \noalign{\smallskip}
             O &   $1.71 \times 10^{-3}$& N$_2$  &   $7.47 \times 10^{-5}$  \\
             C &  $7.26 \times 10^{-4}$  & NH$_3$  &  $7.47 \times 10^{-5}$ \\
             N &  $2.24 \times 10^{-4}$ & S & $3.24 \times 10^{-5}$\\
             H$_2$O  & $4.86 \times 10^{-4}$ & Ar & $7.26 \times 10^{-6}$ \\
             CO$_2$  & $5.31 \times 10^{-4}$ & Kr & $3.39 \times 10^{-9}$ \\
             CO  & $1.77 \times 10^{-4}$  & Xe  & $3.39 \times 10^{-10}$ \\
             CH$_4$  & $1.77 \times 10^{-5}$ \\

            \hline
         \end{tabular}
         \end{center}
         \label{table_AG89}
         \end{table}

We aim to estimate the composition of ices incorporated in Galilean satellites in a way consistent with the formation of Jupiter and its primordial volatile composition, as calculated in A05b. These authors showed that, in order to fit the volatile enrichments measured by the {\it Galileo} probe, only some values of CO$_2$:CO:CH$_4$ and N$_2$:NH$_3$ ratios (consistent with ISM observations of Gibb et al. (2004) and Allamandola et al. (1999)) were allowed. Since solids that were incorporated in the Jovian subnebula initially formed in Jupiter's feeding zone, they shared the same composition than those accreted by proto-Jupiter during its formation. Hence, in our calculations, we adopt the same CO$_2$:CO:CH$_4$ and N$_2$:NH$_3$ ratios in the solar nebula gas-phase as those determined by A05b (see Table \ref{water}).

\subsection{Constraining the abundance of water in Jupiter's feeding zone}

According to the clathrate hydrate trapping theory (Lunine \& Stevenson 1985), the complete clathration of CO, CH$_4$, N$_2$, NH$_3$, H$_2$S, Xe, Kr, and Ar in Jupiter's feeding zone requires an important amount of available crystalline water. This tranlates in a H$_2$O:H$_2$ ratio greater than that deduced from solar gas-phase abundances of elements in the solar nebula (see Table \ref{table_AG89} and Table \ref{water}). This overabundance may result from the inward drift of icy grains (Supulver \& Lin 2000), and from local accumulation of water vapor at radii interior to the water evaporation/condensation front, as described by Cuzzi \& Zahnle (2004). The corresponding minimum molar mixing ratio of water relative to H$_2$ in the solar nebula gas-phase is given by

\begin{equation}
{x_{H_2O} = \sum_{\it{i}} \gamma_i~x_i~\frac{\Sigma(R; T_i,
P_i)_{neb}}{\Sigma(R; T_{H_2O}, P_{H_2O})_{neb}}},
\end{equation}

\noindent where $x_i$ is the molar mixing ratio of the volatile $i$ with respect to H$_2$ in the solar nebula gas-phase, $\gamma_i$ is the required number of water molecules to form the corresponding hydrate or clathrate hydrate (5.75 for a type I clathrate hydrate, 5.66 for a type II clathrate hydrate, 1 for the NH$_3$-H$_2$O hydrate and 0 for CO$_2$ pure condensate), $\Sigma(R; T_i, P_i)_{neb}$ and $\Sigma(R; T_{H_2O}, P_{H_2O})_{neb}$ are the surface density of the nebula at the distance $R$ from the Sun at the epoch of hydratation or clathration of the species $i$ and at the epoch of condensation of water, respectively.\\

Table \ref{water} gives the values of $x_{H_2O}$ in Jupiter's feeding zone, for the  CO$_2$:CO:CH$_4$ and N$_2$:NH$_3$ ratios used in A05b and in this work. Note that, in order to calculate $x_{H_2O}$, we have considered a subsolar abundance for H$_2$S,  similarly to A05b. Indeed, H$_2$S, at the time of its incorporation in icy planetesimals, may have been subsolar in the protoplanetary disk, as a result of the coupling between the oxygen-dependent sulfur chemistry, the FeS kinetics, and the nebular transport processes that affect both oxygen and sulfur abundances (Pasek et al. 2005). Following the calculations described in A05b to fit the observed sulfur enrichment in Jupiter, we have adopted H$_2$S:H$_2$~=~0.60 $\times$ (S:H$_2$)$_\odot$ for N$_2$:NH$_3$~=~10 and H$_2$S:H$_2$~=~0.69 $\times$ (S:H$_2$)$_\odot$ for N$_2$:NH$_3$~=~1 in the solar nebula gas-phase.

\begin{table}[h]
   \caption[]{Calculations of the gas-phase abundance of water $x_{H_2O}$ required to trap all volatile species, except CO$_2$ that condenses as a pure ice, in Jupiter's feeding zone. CO$_2$:CO:CH$_4$ and N$_2$:NH$_3$ gas-phase ratios considered here are those determined by A05b in
the solar nebula gas-phase to fit the enrichments in volatiles in Jupiter.}
  \begin{center}
       \begin{tabular}[]{lcc}
            \hline
            \hline
            \noalign{\smallskip}
               & N$_2$:NH$_3$ = 10  &  N$_2$:NH$_3$ = 1 \\
            \noalign{\smallskip}
             \hline
             \noalign{\smallskip}
            CO$_2$:CO:CH$_4$ = 10:10:1  &  $1.55 \times 10^{-3}$  &  $1.51
\times 10^{-3}$  \\
            CO$_2$:CO:CH$_4$ = 20:10:1  &  -  &  $1.14 \times 10^{-3}$  \\
            CO$_2$:CO:CH$_4$ = 30:10:1  &  -  &  $9.48 \times 10^{-4}$  \\
            CO$_2$:CO:CH$_4$ = 40:10:1  &  -  &  $8.33 \times 10^{-4}$  \\
            \hline
         \end{tabular}
         \end{center}
         \label{water}
         \end{table}

\subsection{Composition of ices incorporated in planetesimals produced in Jupiter's feeding zone}
\label{comp_planetesimaux}

Using the aforementioned water abundances, one can calculate the mass abundances of major volatiles with respect to H$_2$O in icy planetesimals formed in Jupiter's feeding zone. Indeed, the volatile $i$ to water mass ratio in these planetesimals is determined by the relation given by MG04:

\begin{equation}
Y_i = \frac{X_i}{X_{H_2O}} \frac{\Sigma(R; T_i, P_i)_{neb}}{\Sigma(R; T_{H_2O},
P_{H_2O})_{neb}},
\end{equation}

\noindent where $X_i$ and $X_{H_2O}$ are the mass mixing ratios of the volatile $i$ and of H$_2$O with respect to H$_2$ in the solar nebula, respectively. In this calculation, $X_{H_2O}$ is derived from $x_{H_2O}$ in Table \ref{water}.

\section{Turbulent model of the Jovian subnebula}

\subsection{Thermodynamical characteristics of the model}

   The $\alpha$-turbulent model of the Jovian subnebula we considered here is the one proposed in Paper I. The subdisk evolution is divided into two distinct phases. During the first one, the Jovian subnebula is fed by the solar nebula. During this phase, which lasts about 0.56 Myr, the subnebula is in equilibrium since the accretion rate is constant throughout the subdisk. The origin of time has then no influence and is arbitrarily chosen as being the moment when Jupiter has already accreted $\sim 85 \%$ of its total mass. When the solar nebula disappears, the accretion rate at the outer edge of the subdisk decreases to zero, and the subnebula enters its second phase. The subdisk evolves due to the accretion of its own material onto the
planet, and expands outward due to the conservation of angular momentum.

The strategy describing the choice of the different subdisk parameters is given in Paper I and the different parameters of the Jovian subnebula are recalled in Table \ref{table_thermo}.

       \begin{table}[h]
      \caption[]{Thermodynamical parameters of the Jovian subnebula.}
      \begin{center}
         \begin{tabular}[]{lc}
            \hline
            \hline
            \noalign{\smallskip}
               Thermodynamical  &  \\
               parameters  & \\
             \noalign{\smallskip}
             \hline
             \noalign{\smallskip}
             Mean mol. weight (g/mole) &  2.4 \\
             $\alpha$  & $2 \times 10^{-4}$ \\
             Initial disk's radius ($R_{J}$) & 150 \\
             Initial disk's mass ($M_J$) & $3 \times 10^{-3} $ \\
             Initial accretion rate ($M_{J}$/yr) & $9 \times 10^{-7}$ \\
           \noalign{\smallskip}
            \hline
         \end{tabular}
   \end{center}
   \label{table_thermo}
   \end{table}

\subsection{Evolution of volatile rich planetesimals incorporated in the subnebula}

\begin{figure}
   \begin{center}
   \epsfig{file=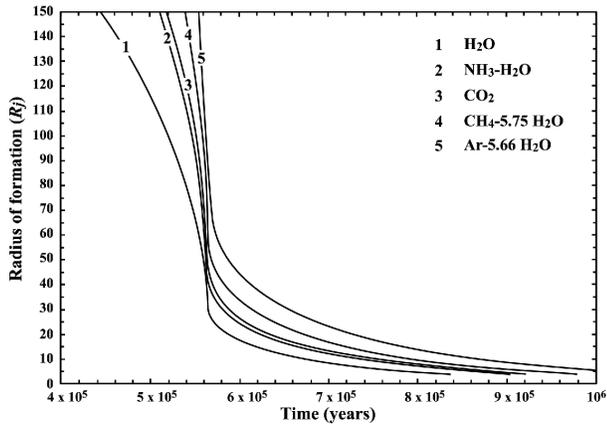,angle=90,width=80mm}
   \end{center}
   \caption{Radii of formation of water ice, NH$_3$-H$_2$O hydrate, CO$_2$ pure condensate, and CH$_4$ and Ar clathrate hydrates in the Jovian subnebula as a function of time. Radii of formation of H$_2$S, Xe, CO, N$_2$ and Kr clathrate hydrates are not represented but are within curves 1 and 5.}
   \label{cond_subneb}
   \end{figure}

   Figure \ref{cond_subneb} illustrates the fate of ices incorporated in planetesimals accreted from the nebula by the subnebula. From this figure, it can be seen that, as soon as they are introduced into the Jovian subnebula, the different ices start to vaporize in the whole subdisk. The different volatile species considered here start to crystallize again in the Jovian subnebula's outer edge between 0.44 and 0.55 Myr. Note that we have considered here the condensation temperatures given in Fig. 1 to calculate the epochs of crystallization of the different ices in the subnebula. Moreover, we did not take into account the ablation due to the friction of planetesimals with gas. Water ice condenses at  $t$~=~0.57 Myr at the orbit of Callisto (26.6 R$_J$) and at $t$~=~0.61 Myr at the orbit of Ganymede (15.1 R$_J$). NH$_3$-H$_2$O hydrate becomes stable at $t$~=~0.59 Myr at the orbit of Callisto and at $t$~=~0.67 Myr at the orbit of Ganymede. CO$_2$ pure condensate is not stable at times earlier than 0.60 Myr at the orbit of Callisto and 0.68 Myr at the orbit of Ganymede. In addition, clathrate hydrates of H$_2$S, Xe, CH$_4$, CO, N$_2$, Kr, and Ar become stable between $t$~=~0.59 Myr and $t$~=~0.68 Myr at the orbit of Callisto, and $t$~=~0.67 Myr and $t$~=~0.78 Myr at the orbit of Ganymede. Icy planetesimals entering into the subnebula at epochs later than those indicated above should keep trapped their volatiles and maintain the Ices/Rocks (I/R) ratios they acquired in the solar nebula. On the other hand, icy planetestimals entering into the subnebula at epochs prior to those determined for preserving ices at the orbits of the two major icy satellites must have lost their content in volatiles in the satellite zone due to vaporization. 

\subsection{Gas-phase chemistry of major C and N bearing volatiles in the subnebula}
\label{gas_chemistry}

 Since ices were all vaporized in the subdisk during at least the first $\sim 0.5$ Myr of the Jovian subnebula evolution, it seems worthwhile to examine the gas-phase reactions that can occur for major C and N volatile species in such an environment.

   Following Prinn \& Fegley (1989), the net reactions relating CO, CH$_4$, CO$_2$, N$_2$ and NH$_3$ in a gas dominated by H$_2$ are

\begin{equation}
\mathrm{CO + H_2O = CO_2 +H_2}
\label{eq_chim1}
\end{equation}

\begin{equation}
\mathrm{CO + 3H_2 = CH_4 +H_2O}
\label{eq_chim2}
\end{equation}

\begin{equation}
\mathrm{N_2 + 3H_2 = 2NH_3}
\label{eq_chim3}
\end{equation}

\noindent which all proceed to the right with decreasing temperature at constant pressure. Reaction (\ref{eq_chim1}) has been recently studied by Talbi \& Herbst (2002) who demonstrated that its rate coefficient is negligible, even at temperature as high as 2000 K (of the order of $\sim 4.2~\times~10^{-22}$~cm$^3$~s$^{-1}$). Such a high temperature range is only reached at distances quite close to Jupiter and at early epochs in the Jovian subnebula (see Fig. 6 in Paper I). As a result, the amount of carbon species produced through this reaction is insignificant during the whole lifetime of the subnebula.

\begin{figure}
   \begin{center}
   \epsfig{file=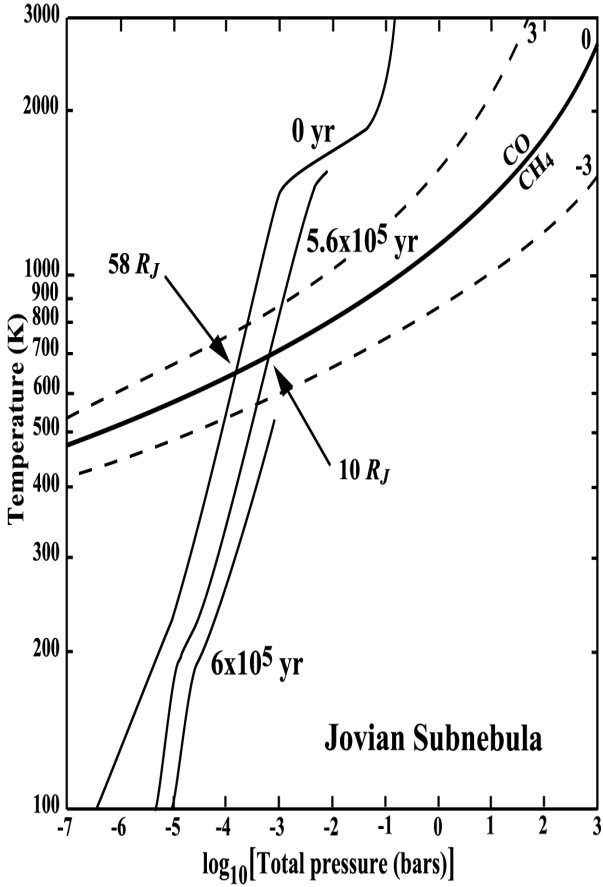,height=120mm,width=80mm}
   \end{center}
   \caption{Calculated ratios of CO:CH$_4$ in the Jovian subnebula at the equilibrium. The solid line labelled CO-CH$_4$ corresponds to the case where the abundances of the two gases are equal. When moving towards the left side of the solid line, CO/CH$_4$ increases, while moving towards the right side of the solid line, CO/CH$_4$ decreases. The dotted contours labelled -3, 0, 3 correspond to log$_{10}$ CO:CH$_4$ contours. Thermodynamical conditions in our evolutionary turbulent model of the Jovian subdisk are represented at three epochs of the subnebula. The Jovianocentric distance, in $R_J$, is indicated by arrows when CO:CH$_4$ = 1 for $t$~=~0~and~0.56 Myr (transition epoch between the two phases of the subnebula evolution). }
   \label{CO-CH4_eq}
   \end{figure}

\begin{figure}
   \begin{center}
   \epsfig{file=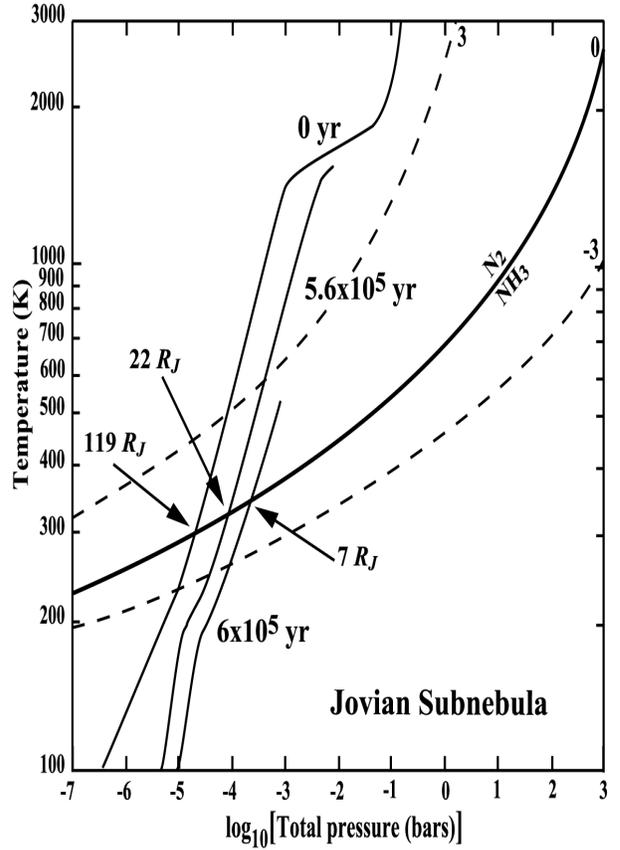,height=115mm,width=80mm}
   \end{center}
   \caption{Same as Fig. \ref{CO-CH4_eq}, but for calculated ratios of N$_2$:NH$_3$ at the equilibrium. The Jovianocentric distance, in $R_J$, is indicated by arrows when N$_2$:NH$_3$ = 1 for $t$ = 0, 0.56 Myr and 0.6 Myr of our turbulent model.}
   \label{N2-NH3_eq}
   \end{figure}

Reactions (\ref{eq_chim2}) and (\ref{eq_chim3}) are respectively illustrated by Figs. \ref{CO-CH4_eq} and \ref{N2-NH3_eq}. The calculations are performed using the method described in Mousis et al. (2002a) where the reader is referred for details. At the equilibrium, CO:CH$_4$ and N$_2$:NH$_3$ ratios depend only upon local conditions of temperature and pressure (Prinn \& Barshay 1977; Lewis \& Prinn 1980; Smith 1998). CO:CH$_4$ and N$_2$:NH$_3$ ratios of 1000, 1, and 0.001 are plotted in Figs. \ref{CO-CH4_eq} and \ref{N2-NH3_eq}, and compared to our turbulent model at 
three different epochs (0 yr, 0.56 Myr and 0.6 Myr). These figures show that, when kinetics of chemical reactions are not considered, CH$_4$ and NH$_3$ progressively dominate with time in the major part of our turbulent model of the Jovian subnebula rather than CO and N$_2$. 

\begin{figure}
   \begin{center}
   \epsfig{file=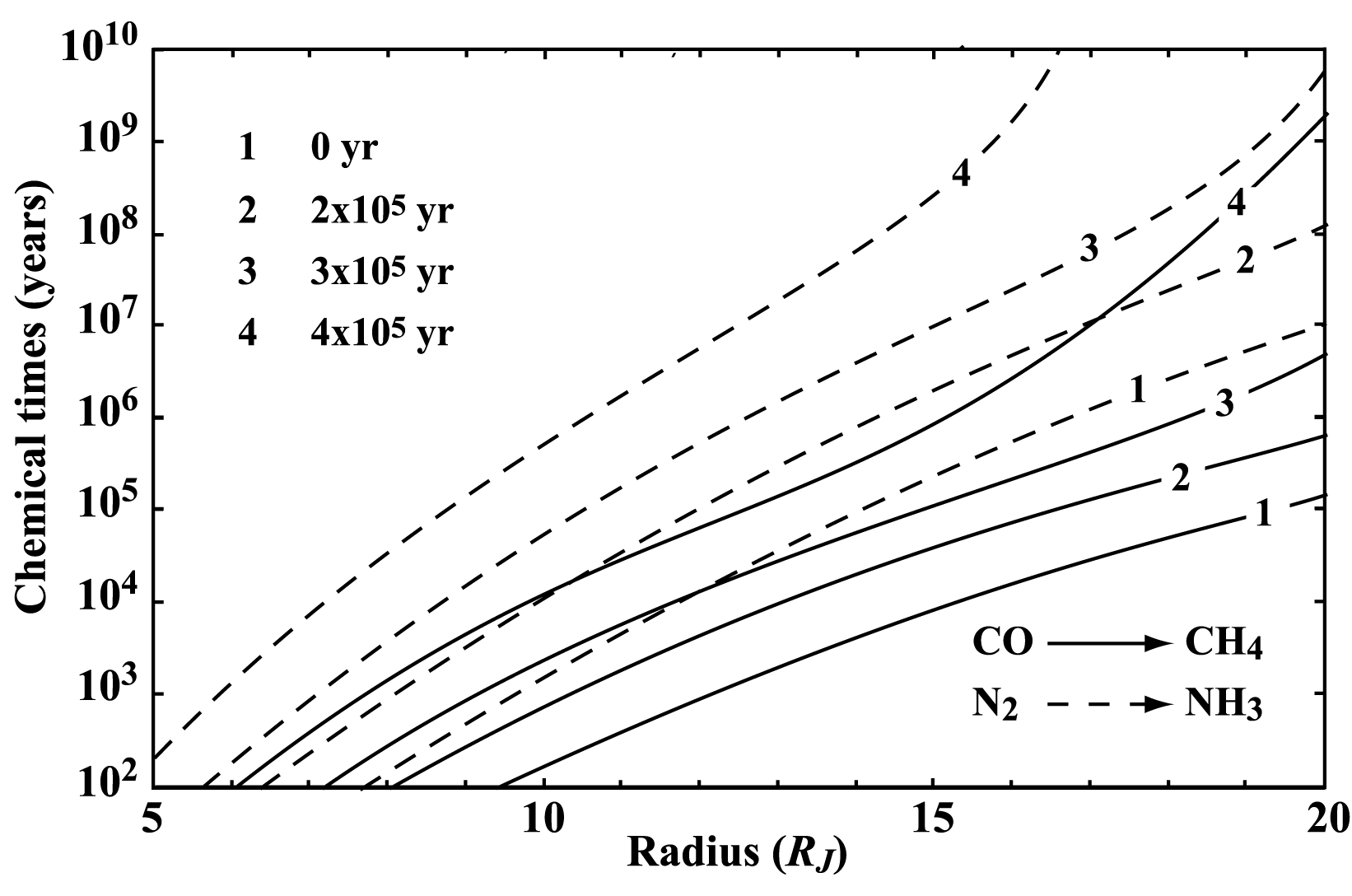,height=65mm,angle=-90,width=85mm}
   \end{center}
   \caption{Chemical times profiles calculated for CO:CH$_4$ and N$_2$:NH$_3$ conversions in our model of the Jovian subnebula. The conversion of CO to CH$_4$ and of N$_2$ to NH$_3$ is fully inhibited, except quite close to Jupiter and at the early times of the subnebula evolution.}
   \label{tps_chim}
   \end{figure}

However, the actual CO:CH$_4$ and N$_2$:NH$_3$ ratios depend on the chemical timescales, which characterize the rates of CO to CH$_4$ and N$_2$ to NH$_3$ conversions in our model of the Jovian subnebula. We have calculated these chemical times from the data given by Prinn \& Barshay (1977), Lewis \& Prinn (1980), and Smith (1998), and using the temperature and pressure profiles derived from our turbulent model. The results, calculated at several epochs of the subnebula's life and at different distances to Jupiter, are represented in Fig. \ref{tps_chim}. Taking into account the kinetics of chemical reactions, one can infer that the efficiency of the  conversion is limited only to the inner part of the Jovian subnebula and at early times of its first phase. This implies that CO:CH$_4$ and N$_2$:NH$_3$ ratios remain almost constant during the whole lifetime of the Jovian subnebula. Moreover, since reaction (\ref{eq_chim1}) plays no role in the Jovian subnebula, the CO$_2$:CO ratio also remains fixed during its lifetime.\\

Finally, these conclusions are compatible with those found by MG04 for their colder subnebula model and imply that the CO$_2$:CO:CH$_4$ and N$_2$:NH$_3$ ratios in the subnebula gas-phase were close to the values acquired in Jupiter's feeding zone once these species were trapped or
condensed. From these initial gas-phase conditions in the Jovian subnebula, it is now possible to examine the composition of regular satellites ices if these bodies formed from planetesimals produced in this environment.

\section{Constraining the composition of ices incorporated in regular icy satellites}

Following Paper I, the forming protosatellites migrate inwards the Jovian subnebula under the influence of type I migration (Ward 1997). Since protosatellites are susceptible to migrate at different epochs of the Jovian subnebula's life, two opposite regular satellite formation scenarios can then be derived. In the first one, regular icy satellites of Jupiter have been accreted from planetesimals that were preserved from vaporization when they entered into the subnebula after several hundreds of thousands years of existence. In the second scenario, regular icy satellites have been accreted from icy planetesimals that formed in the subnebula. We now explore the consequences of these two scenarios on the resulting composition of ices incorporated in the Jovian regular icy satellites.

\subsection{First scenario: icy planetesimals produced in the solar nebula}

The hypothesis of satellite formation from primordial planetesimals (i.e. planetesimals that were produced in Jupiter's feeding zone without subsequent vaporization) is supported by the recent work of A05b who found, with their nominal model for interpreting the enrichments in volatiles in Jupiter's atmosphere, that I/R in solids accreted by the giant planet is similar to that estimated in the current Ganymede and Callisto. In this scenario, the mass abundance of major volatiles with respect to H$_2$O in the Jovian regular icy satellites is equal to that in planetesimals formed in Jupiter's feeding zone, and calculated in Sect. \ref{comp_planetesimaux}.

\subsection{Second scenario: icy planetesimals produced in the Jovian subnebula}

From Fig. \ref{cond_subneb}, it can be seen that ices entering into the Jovian subdisk were all vaporized at epochs prior to $\sim 0.5$ Myr. With time, the subnebula cooled down and volatiles started to crystallize again following the same condensation sequence as that described in the solar nebula (see Fig. \ref{cool_curve}). One can link the resulting volatile $i$ to water mass ratio $(Y_i)_{sub}$ in solids formed into the Jovian subnebula to the initial one $(Y_i)_{feed}$ in planetesimals produced in Jupiter's feeding zone through the following relation:

\begin{equation}
(Y_i)_{sub}~=~f_i \times~(Y_i)_{feed},
\end{equation}

\noindent where $f_i$ is the fractionation factor due to the consecutive vaporization and condensation of volatile $i$ in the subdisk. The fractionation factor $f_i$ is given by:

\begin{equation}
f_i = \frac{\Sigma(R; T_i, P_i)_{sub}}{\Sigma(R; T_{H_2O}, P_{H_2O})_{sub}},
\end{equation}

\noindent where $\Sigma(R; T_i, P_i)_{sub}$ and $\Sigma(R; T_{H_2O}, P_{H_2O})_{sub}$ are the surface densities in the Jovian subnebula, at the distance $R$ from Jupiter, and at the epochs of trapping of species $i$ and of H$_2$O condensation, respectively. Using condensation temperatures of the different ices formed in the subnebula similar to those calculated in Jupiter's feeding zone, $f_i$ remains almost constant (10 $\%$ variations at most) in the whole subdisk. Values of $f_i$ range between 0.40 and 0.76 and are given for each species in Table \ref{table_fi}. 

In summary, if regular icy satellites were accreted from solids produced in the subnebula, the resulting volatile $i$ to water mass ratio $(Y_i)_{sub}$ in their ices is that estimated in planetesimals formed in Jupiter's feeding zone (see Sect. \ref{comp_planetesimaux}) multiplied by the fractionation factor $f_i$.

\begin{table}[h]
   \caption[]{Mean values of the fractionation factor $f_i$ calculated for icy
planetesimals 
produced in the Jovian subnebula.}
 \begin{center}
       \begin{tabular}[]{lclc}
            \hline
            \hline
            \noalign{\smallskip}
            Species & $f_i$  &  Species  & $f_i$\\
            \noalign{\smallskip}
             \hline
             \noalign{\smallskip}
             NH$_3$:H$_2$O  &  0.76    &  H$_2$S:H$_2$O   &  0.74 \\
             CO$_2$:H$_2$    &  0.71    &  CH$_4$:H$_2$O   &  0.57 \\
             Xe:H$_2$O         &  0.59    &   CO:H$_2$O      &  0.49 \\
             N$_2$:H$_2$      &  0.47    &  Kr:H$_2$O        &  0.45\\
             Ar:H$_2$O        &  0.40 \\
            \hline
         \end{tabular}
         \end{center}
         \label{table_fi}
         \end{table}

\subsection{Composition of regular satellites ices}

Table \ref{table_comp}  summarizes the composition range that can be found in the Jovian regular satellites ices formed in the framework of the first scenario and assuming that most of the trapped volatiles were not lost during the accretion and the thermal history of the satellites. From this table, it can be seen that CO$_2$:H$_2$O, CO:H$_2$O and CH$_4$:H$_2$O mass ratios vary between $3.7 \times 10^{-1}$ and $1.15$, between $1.3 \times 10^{-1}$ and $1.8 \times 10^{-1}$, and between $9 \times 10^{-3}$ and $1.1 \times 10^{-2}$, respectively, in the interiors of regular icy satellites, as a function of CO$_2$:CO:CH$_4$ and N$_2$:NH$_3$ gas-phase ratios assumed in the solar nebula. Similarly, N$_2$:H$_2$O, NH$_3$:H$_2$O and H$_2$S:H$_2$O ratios should be between $3.8 \times 10^{-2}$ and $6.9 \times 10^{-2}$, between $5 \times 10^{-3}$ and $6.2 \times 10^{-2}$, and between $1.9 \times 10^{-2}$ and $3.6 \times 10^{-2}$, respectively. Low amounts of Ar, Kr, and Xe should also exist in the interiors of regular icy satellites. In the second scenario, the resulting 
volatile $i$ to water mass ratio in regular icy satellites must be revised down compared to the values quoted above and in Table \ref{table_comp}, using the fractionation factors given  in Table \ref{table_fi}.

\begin{table*}
   \caption[]{Calculations of the ratios of trapped masses of volatiles to the mass of H$_2$O ice in regular icy satellites accreted from planetesimals formed in Jupiter's feeding zone. Gas-phase abundances of H$_2$O are given in Table 2 and gas-phase abundances of elements, except S (see text), are assumed to be solar (see Table 1). Ranges of CO$_2$:CO:CH$_4$ and N$_2$:NH$_3$ gas-phase ratios considered here are those determined by A05b in the solar nebula gas-phase to fit the enrichments in volatiles in Jupiter (see text).}
  \begin{center}
       \begin{tabular}[]{lccccc}
            \hline
            \hline
            \noalign{\smallskip}
            & N$_2$:NH$_3$ = 10 & \multicolumn{4}{c}{N$_2$:NH$_3$ = 1}
\\
            \noalign{\smallskip}
            Species  & CO$_2$:CO = 1  & CO$_2$:CO = 1  & CO$_2$:CO =
2  & CO$_2$:CO = 3  & CO$_2$:CO = 4   \\
            \noalign{\smallskip}
             \hline
             \noalign{\smallskip}
             CO$_2$:H$_2$O      &  $3.72 \times 10^{-1}$   &  $3.84 \times 10^{-1}$  & $6.92 \times 10^{-1}$   &  $9.43 \times 10^{-1}$  &  $1.15$ \\
             CO:H$_2$O               &  $1.78 \times 10^{-1}$   &  $1.83 \times 10^{-1}$  &  $1.63 \times 10^{-1}$  &  $1.47 \times 10^{-1}$  & $1.34 \times 10^{-1}$ \\
             CH$_4$:H$_2$O      &  $1.14 \times 10^{-2}$   &  $1.17 \times 10^{-2}$  &  $1.04 \times 10^{-2}$  &  $9.44 \times 10^{-3}$  & $8.60 \times 10^{-3}$ \\
             N$_2$:H$_2$O       &  $5.34 \times 10^{-2}$   &  $3.82 \times 10^{-2}$  & $5.06 \times 10^{-2}$   &  $6.07 \times 10^{-2}$  & $6.90 \times 10^{-2}$ \\
             NH$_3$:H$_2$O   &  $4.60 \times 10^{-3}$   &  $3.43 \times 10^{-2}$  &  $4.55 \times 10^{-2}$  &  $5.45 \times 10^{-2}$  & $6.20 \times 10^{-2}$  \\
             H$_2$S:H$_2$O      &  $1.93 \times 10^{-2}$   &  $1.99 \times 10^{-2}$  &  $2.63 \times 10^{-2}$  &  $3.16 \times 10^{-2}$  & $3.59 \times 10^{-2}$ \\
             Ar:H$_2$O                  &  $4.17 \times 10^{-3}$   &  $4.29 \times 10^{-3}$  &  $5.69 \times 10^{-3}$  &  $6.83 \times 10^{-3}$  & $7.76 \times 10^{-3}$ \\
             Kr:H$_2$O                  &  $4.89 \times 10^{-6}$   &  $5.04 \times 10^{-6}$  &  $6.68 \times 10^{-6}$  &  $8.01 \times 10^{-6}$  & $9.11 \times 10^{-6}$  \\
             Xe:H$_2$O                  &  $9.29 \times 10^{-7}$   &  $9.57 \times 10^{-7}$  &  $1.27 \times 10^{-6}$  &  $1.52 \times 10^{-6}$  & $1.73 \times 10^{-6}$  \\
            \hline
         \end{tabular}
         \end{center}
         \label{table_comp}
         \end{table*}

\section{Summary and discussion}

In this work, we have used the evolutionary turbulent model of the Jovian subnebula described in Paper I to calculate the composition of ices incorporated in the regular icy satellites of Jupiter. The model of the Jovian subnebula we used here evolves in two distinct phases during its lifetime. In the first phase, the Jovian subnebula is fed by the solar nebula as long as the latter has not been dissipated. In the second phase, the solar nebula has disappeared and the subnebula progressively empties by accreting its material onto the forming Jupiter. Solids entering into the Jovian subnebula and that may ultimately lead to the Jovian satellite formation are assumed to have been produced in the feeding zone of proto-Jupiter prior to its appearance. Some of these solids were coupled with the material flowing in the Jovian subnebula from the solar nebula during the first phase of its evolution, due to their submeter dimensions, while larger of them, with heliocentric orbits, may have been captured by the subdisk when they came through. 

We have considered CO$_2$, CO, CH$_4$, N$_2$, NH$_3$, H$_2$S, Ar, Kr, and Xe as the major volatile species existing in the gas-phase of Jupiter's feeding zone. All these volatiles, except CO$_2$, have been trapped under the form of hydrates or clathrate hydrates in Jupiter's feeding zone during the cooling of the solar nebula. CO$_2$ crystallized as a pure condensate prior to be trapped by water and formed the only existing condensed form of CO$_2$ in the feeding zone of Jupiter.

We employed CO$_2$:CO:CH$_4$ and N$_2$:NH$_3$ ratios consistent with those used by A05b, namely CO$_2$:CO:CH$_4$ between 10:10:1 and 40:10:1, and N$_2$:NH$_3$ between 1 and 10 in the gas-phase of Jupiter's feeding zone. Such a range of values is compatible with those measured in ISM or estimated in the solar nebula. This allowed us to determine the corresponding minimum H$_2$O:H$_2$ gas-phase ratios required to trap all volatiles (except CO$_2$) in the giant planet's feeding zone.

Moreover, since, according to our model, ices contained in solids entering into the subnebula before $\sim 0.5$ Myr were all vaporized, we have followed the net gas-phase chemical reactions relating CO, CH$_4$, CO$_2$, N$_2$, and NH$_3$ in this environment.  We then concluded that these reactions are mostly inefficient in the Jovian subnebula, in agreement with the previous work of MG04. This involves that CO$_2$:CO:CH$_4$ and N$_2$:NH$_3$ ratios were not essentially different from those acquired in the feeding zone of Jupiter, once these species were trapped or condensed in the subdisk. In addition, in order to estimate the mass abundances of the major volatile species with respect to H$_2$O in the interiors of the Jovian regular icy satellites, we considered the formation of these bodies by following two opposite scenarios.

In the first scenario, regular icy satellites were accreted from planetesimals that have been preserved from vaporization during their migration in the Jovian subnebula. This assumption is in agreement with the work of A05b who found, with their nominal model for interpreting the enrichments in volatiles in Jupiter's atmosphere (N$_2$:NH$_3$~=~1 and CO$_2$:CO:CH$_4$~=~30:10:1 in the solar nebula gas-phase), that I/R in planetesimals accreted by the giant planet is similar to those estimated by Sohl et al. (2002) in Ganymede and Callisto. This allowed us to estimate the ratios of the trapped masses of volatiles to the mass of H$_2$O ice in the regular icy satellites, assuming that these species were not lost during their accretion and their thermal history. 

In the second scenario, regular icy satellites were accreted from planetesimals produced in the subnebula. Indeed, in the framework of our model, ices contained in solids were entirely vaporized if they entered at early epochs into the Jovian subdisk. With time, the subnebula cooled down and ices crystallized again in the subnebula prior to having been subsequently incorporated into the growing planetesimals. In this second scenario, assuming, as in the first one, that the regular icy satellites did not lose volatiles during their accretion phase and their thermal history, we have also estimated the composition range of ices trapped in their interiors. In that scenario, the amount of ices incorporated in regular icy satellites should be lower than in the previous one where planetesimals were produced in the solar nebula.

In both scenarios, the calculated composition of the Jovian regular satellites ices is consistent with some evidences of carbon and nitrogen volatile species in these bodies, even if the presence of some predicted components has yet to be verified. For example, reflectance spectra returned by the Near-Infrared Mapping Spectrometer (NIMS) aboard the Galileo spacecraft revealed the presence of CO$_2$ over most of Callisto's surface (Hibbitts et al. 2000). Moreover, Hibbitts et al. (2002) suggested that CO$_2$ would be contained in clathrate hydrates located in the subsurface of Callisto and would be stable over geologic time unless exposed to the surface. CO$_2$ has also been detected on the surface of Ganymede (Hibbitts et al. 2003). In addition, one explanation for the internal magnetic fields discovered in both Ganymede and Callisto (Kivelson et al. 1997, 1999) invokes the presence of subsurface oceans within these satellites (Sohl et al. 2002). The presence of such deep oceans is probably linked to the presence of NH$_3$ in the interiors of these satellites, since this component decreases the solidus temperature by several tens of degrees (Mousis et al. 2002b; Spohn \& Schubert 2003).

Subsequent observations are required to determine which of both presented formation scenarios is the most realistic. On the basis of isotopic exchanges calculations between HDO and H$_2$ in the solar nebula, Mousis (2004) estimated that the D:H ratio in the Jovian regular icy satellites is between $\sim$ 4 and 5 times the solar value, assuming they were formed from planetesimals produced in Jupiter's feeding zone. On the other hand, icy satellites formed from planetesimals produced in the Jovian subnebula should present a lower D:H ratio in H$_2$O ice since an additional isotopic exchange occurred between HDO and H$_2$ in the subdisk gas-phase. Such estimates, compared with further $in$ $situ$ measurements of the D:H ratio in H$_2$O ice on the surfaces of the Jovian regular satellites, should allow to check the validity of the two proposed formation scenarios.

\begin{acknowledgements}
This work was supported in part by the Swiss National Science Foundation. OM was partly supported by an ESA external fellowship, and this support is gratefully acknowledged. We thank the referee for useful comments on the manuscript.
\end{acknowledgements}

{}


\begin{thebibliography}{}

\bibitem[Alibert2005]{Alibert2005a} Alibert, Y., Mousis, O., \& Benz, W. \ 2005a, A\&A, 439, 1205 (Paper I)
\bibitem[Alibert2005]{Alibert2005b} Alibert, Y., Mousis, O., \& Benz, W. \ 2005b, ApJL, 622, 145 (A05b)
\bibitem[Alibert2005]{Alibert2005c} Alibert, Y., Mousis, O., Mordasini, C., et al. \ 2005c, ApJL, 626, 57 
\bibitem[1989]{anders} Anders, E., \& Grevesse, N. 1989, Geochim. Cosmochim. Acta, 53, 197
\bibitem[1999]{allamandola} Allamandola, L.~J., Bernstein, M.~P., Sandford, S.~A., et al. \ 1999, Space Sci. Rev., 90, 219
\bibitem[Chick \& Cassen 1997]{Chick1997} Chick, K.~M., \& Cassen, P.~\ 1997, ApJ, 477, 398
\bibitem[Cuzzi \& Zahnle 2004]{Cuzzi2004} Cuzzi, J.~N., \& Zahnle, K.~J.~\ 2004, ApJ, 614, 490
\bibitem[Drouart2000]{Drouart2000} Drouart, A., Dubrulle, B., Gautier, D. ~et al.~\ 1999, Icarus, 140, 129
\bibitem[Fegley 2000]{Fegley 2000} Fegley, B.~Jr. \ 2000, Space Sci. Rev., 92, 177
\bibitem[Gibb et al. 2004]{Gibb et al. 2004} Gibb, E.~L., Whittet, D.~C.~B., Boogert, A.~C.~A., ~et al.~\ 2004, ApJS, 151, 35
\bibitem[Hersant et al. 2001]{Hersant et al. 2001} Hersant, F., Gautier, D. \& Hur{\' e}, J.-M. ~\ 2001, \apj, 554, 391
\bibitem[Hibbitts et al. 2000]{Hibbitts et al. 2000} Hibbitts, C.~A., Mc Cord, T.~B., Hansen, J.  ~et al.~\ 2000, \jgr, 105, 22541
\bibitem[Hibbitts et al. 2002]{Hibbitts et al. 2002} Hibbitts, C.~A., Klemaszewski, J.~E., Mc Cord, T.~B., ~et al.~\ 2002, \jgr, 107, 14-1
\bibitem[Hibbitts et al. 2003]{Hibbitts et al. 2003} Hibbitts, C.~A., Pappalardo, R.~T., Hansen, G.~B. ~et al.~\ 2003, \jgr, 108, 2-1
\bibitem[Kivelson et al. 1997]{Kivelson et al. 1997} Kivelson, M.~G., Khurana, K.~K., Coroniti, F.~V., ~et al.~\ 1997, \grl, 24, 2155
\bibitem[Kivelson et al. 1999]{Kivelson et al. 1999} Kivelson, M.~G., Khurana, K.~K., Stevenson, D.~J., ~et al.~\ 1999, \jgr, 104, 4609
\bibitem[Kouchi et al. 1994]{Kouchi et al. 1994} Kouchi, A., Yamamoto, T, Kozasa, T.,  ~et al.~\ 1994, \aap, 290, 1009
\bibitem[Lewis \& Prinn 1980]{Lewis1980} Lewis, J.~S., \& Prinn, R.~G. \ 1980, ApJ, 238, 357
\bibitem[1999]{handbook} Lide, D. R. 1999, CRC Handbook of Chemistry and Physics, ed .Lide, D. R. (Boca Raton: CRC press LLC), 6-59
\bibitem[Lunine \& Stevenson(1985)]{1985ApJS...58..493L} Lunine, J.~I., \& Stevenson, D.~J.\ 1985, \apjs, 58, 493
\bibitem[McCord et al. 1998]{McCord et al. 1998} McCord, T.~B., Hansen, G.~B., Clark, R.~N., ~et al.~\ 1998, \jgr, 103, 8603
\bibitem[Mousis et al. 2000]{Mousis2000} Mousis, O., Gautier, D., Bockel{\' e}e-Morvan, D., ~et al. \ 2000, Icarus, 148, 513
\bibitem[Mousis et al. 2002a]{Mousis2002a} Mousis, O., Gautier, D., \& Bockel{\' e}e-Morvan, D. \ 2002a, Icarus, 156, 162
\bibitem[Mousis et al. 2002b]{Mousis2002b} Mousis, O., Pargamin, J., Grasset, O. ~et al.~\ 2002b, \grl, 29, 2192
\bibitem[Mousis 2004]{Mousis 2004} Mousis, O. \ 2004, A\&A, 414, 1165
\bibitem[Mousis \& Gautier. 2004]{Mousis and Gautier 2004} Mousis, O., \& Gautier, D. \ 2004, \planss, 52, 361
\bibitem[Owen 1999]{Owen et al. 1999} Owen, T.~C., Mahaffy, P.~R., Niemann, H.~B., ~et al.~\  1999, Nature, 402, 269
\bibitem[Pasek et al. 2005]{Pasek et al. 2005} Pasek, M.~A., Milsom, J.~A., Ciesla, F.~J. ~et al.~\ 2005, Icarus, in press
\bibitem[Prinn \& Barshay 1977]{Prinn and Barshay 1977} Prinn, R.~G., \& Barshay, S.~S. \ 1977, Science, 198, 1031
\bibitem[Prinn \& Fegley 1989] {Prinn and Fegley 1989} Prinn, R.~G., \& Fegley Jr.,~B. \ 1989, In Origin and Evolution of Planetary and Satellites Atmospheres (Atreya, S.~K., Pollack, J.~B., Matthews, M.~S., Eds), The University of Arizona Press, Tucson, 78
\bibitem[Saumon \& Guillot 2004]{Saumon and Guillot 2004} Saumon, D., \& Guillot, T. \ 2004, \apj, 609, 1170
\bibitem[Smith 1998]{Smith 1998} Smith, M.~D. \ 1998, Icarus, 132, 176
\bibitem[Sohl2002]{Sohl et al. 2002} Sohl, F., Spohn, T., Breuer, D., ~et al.~\ 2002, Icarus, 157, 104
\bibitem[Spohn \& Schubert 2003]{Spohn2003} Spohn, T. \& Schubert, G. \ 2003, Icarus, 161, 456
\bibitem[2000]{supulver} Supulver, K. D., \& Lin, D. N. C. 2000, Icarus, 146, 525
\bibitem[Talbi \& Herbst 2002]{Talbi and Herbst 2002} Talbi, D., \& Herbst, E. \ 2002, A\&A, 386, 1139
\bibitem[Ward 1997]{Ward 1997} Ward, W.~R.  1997, ApJ, 482, L211


\end{thebibliography}
\end{document}